\documentclass[12pt]{article}
\usepackage{a4p}
\usepackage{epsfig}
\usepackage{amssymb}
\usepackage{amsmath}
\usepackage{cite}
\newcommand{\aow} {\mbox{$a_{0}^{\mathrm{W}}$}}
\newcommand{\aoz} {\mbox{$a_{0}^{\mathrm{Z}}$}}
\newcommand{\aov} {\mbox{$a_{0}^{\mathrm{V}}$}}
\newcommand{\acw} {\mbox{$a_{\mathrm{c}}^{\mathrm{W}}$}}
\newcommand{\acz} {\mbox{$a_{\mathrm{c}}^{\mathrm{Z}}$}}
\newcommand{\acv} {\mbox{$a_{\mathrm{c}}^{\mathrm{V}}$}}

\newcommand{\PWp} {\mbox{$\mathrm{W^+}$}}
\newcommand{\PWm} {\mbox{$\mathrm{W^-}$}}
\newcommand{\Pgg} {\mbox{$\gamma$}}
\newcommand{\Pep}{\mbox{$\mathrm{e^+}$}}
\newcommand{\Pem}{\mbox{$\mathrm{e^-}$}}
\newcommand{\ipb}{\mbox{$\mathrm{pb}^{-1}$}}
\newcommand{\sigqqgg}{\mbox{${\sigma_{\mathrm{q\overline{q}}\gamma\gamma}}$}}
\newcommand{\WW}{\PWp\PWm}

\newcommand{\qq}{\mbox{$\mathrm{q\overline{q}}$}}

\newcommand{\roots}{\mbox{$\sqrt{s}$}}
\newcommand{\Eg}{E_{\gamma}}

\newcommand{\actg}{|\cos\theta_{\gamma}|}

\newcommand{\mrec}{M_{\mathrm{rec}}}

\newcommand{\nngg}{\nu\overline{\nu}\Pgg\Pgg}

\newcommand{\qqgg}{\mathrm{q\bar{q}\gamma\gamma}}

\newcommand{\effqqgg}
{\mbox{$\varepsilon_{{\mathrm{q\overline{q}}}\gamma\gamma}$}}

\newcommand{\cosgi}{\mbox{$\cos\theta_{\gamma i}$}}
\newcommand{\cosgo}{\mbox{$\cos\theta_{\gamma 1}$}}
\newcommand{\cosgt}{\mbox{$\cos\theta_{\gamma 2}$}}

\newcommand{\thegj}{\mbox{$\theta_{\gamma-\mathrm{JET}}$}}
\newcommand{\cosgqi}{\mbox{$\cos\theta^i_{\gamma\mathrm{q}}$}}
\newcommand{\thegqi}{\mbox{$\theta^i_{\gamma\mathrm{q}}$}}
\newcommand{\thegami}{\mbox{$\theta_{\gamma i}$}}
\newcommand{\Egamo}{\mbox{$E_{\gamma 1}$}}
\newcommand{\Egamt}{\mbox{$E_{\gamma 2}$}}
\newcommand{\Egami}{\mbox{$E_{\gamma i}$}}

\newcommand{\epem}{\mbox{$\mathrm{e^+ e^-}$}}

\newcommand{\mpmm}{\mbox{$\mu^+\mu^-$}}

\newcommand{\tptm}{\mbox{$\tau^+\tau^-$}}

\newcommand{\Zzero}{\mbox{${\mathrm{Z}}$}}
\newcommand{\Zgamma}{\mbox{${\mathrm{Z}} / \gamma$}}

\newcommand{\WWg}{\mbox{\WW$\gamma$}}

\newcommand{\nback}{\mbox{$N^{\mathrm{MC}}_{\mathrm{back}}$}}

\newcommand{\ZZ}{\mbox{\Zzero\Zzero}}

\newcommand{\qqg}{\mbox{$\mathrm{q\overline{q}}\gamma$}}

\newcommand{\Mz}{\mbox{$M_{\mathrm{Z}}$}}

\newcommand{\Mqq}{\mbox{$M_{\mathrm{q}\overline{\mathrm{q}}}$}}

\newcommand{\Gz}{\mbox{$\Gamma_{\mathrm{Z}}$}}

\newcommand{\Egam}{\mbox{$E_\gamma$}}
\newcommand{\Ebeam}{\mbox{$E_{\mathrm{beam}}$}}
\newcommand{\Opal}{\mbox{OPAL}}

\newcommand{\cosg}{\mbox{$\cos\theta_\gamma$}}

\newcommand{\cosgq}{\mbox{$\cos\theta_{\gamma\mathrm{q}}$}}
\newcommand{\cosgj}{\mbox{$\cos\theta_{\gamma-\mathrm{JET}}$}}
\newcommand{\GeV}{\mbox{$\mathrm{GeV}$}}

\newcommand{\etal}     {{\it et al}.,\,\ }

\newcommand{\PLB}[3]  {Phys.\ Lett.\ {\bf B#1} (#2) #3}
\newcommand{\ZPC}[3]  {Z.\ Phys.\ {\bf C#1} (#2) #3}
\newcommand{\EPJC}[3]  {Eur.\ Phys.\ J.\ {\bf C#1} (#2) #3}
\newcommand{\JPHYSG}[3]  {J.\ Phys.\ {\bf G#1} (#2) #3}
\newcommand{\NIMA}[3] {Nucl.\ Instr.\ Meth.\ {\bf A#1} (#2) #3}

\newcommand{\EP}[1]   {CERN-EP/{#1}}

\newcommand{\CPC}[3]  {Comp.\ Phys.\ Comm.\ {\bf #1} (#2) #3}
\def\opalabbiendi{OPAL Collaboration, G.\ Abbiendi \etal}

\def\opalahmet{OPAL Collaboration, K.\ Ahmet \etal}

\newcommand{\nunugpv}{NUNUGPV}
\newcommand{\NUNUGPV}{NUNUGPV}

\newcommand{\KoralW}{\mbox{KORALW}}

\newcommand{\kkff}{\mbox{KK2F}}
\newcommand{\KKFF}{\mbox{KK2F}}

\def\etal{\mbox{{\it et al.}}}
\begin{document}

\begin{titlepage}

\begin{center}
{\Large EUROPEAN ORGANISATION FOR NUCLEAR RESEARCH}
\end{center}

\begin{flushright}
    CERN-PH-EP/2004-003\\
    21st January 2004
\end{flushright}
\bigskip
\begin{center}
    \Large\bf\boldmath
Constraints on Anomalous Quartic Gauge Boson Couplings from $\nngg$ 
and $\qqgg$ Events at LEP2
\end{center}
\bigskip
\bigskip
%
%
\begin{center}
\large 
The OPAL Collaboration \\
\bigskip
\end{center}
%
%
\begin{abstract}

Anomalous quartic couplings between the electroweak gauge bosons may contribute 
to the $\nngg$ and $\qqgg$ final states produced in e$^+$e$^-$ collisions. This 
analysis uses the LEP2 OPAL data sample at centre-of-mass energies up to 209~GeV.
Event selections identify $\nngg$ and $\qqgg$ events in which the two photons are 
reconstructed within the detector acceptance. The cross-section for the process
$\epem\rightarrow\qqgg$ is measured. Averaging over all energies, the ratio of the
observed $\epem\rightarrow\qqgg$ cross-section to the Standard Model expectation is
\begin{eqnarray*}
    \mathrm{R}(\mathrm{data}/\mathrm{SM}) = 0.92 \pm 0.07 \pm 0.04, 
\end{eqnarray*}
where the errors represent the statistical and systematic uncertainties
respectively.
The $\nngg$ and $\qqgg$ data are used to constrain possible anomalous 
$\WW\gamma\gamma$ and $\ZZ\gamma\gamma$ couplings. Combining with previous OPAL 
results from the $\WWg$ final state, the 95\,\% confidence level limits on the 
anomalous coupling parameters $\aoz$, $\acz$, $\aow$ and $\acw$ are found to be:
\begin{eqnarray*}
 -0.007~\mathrm{GeV}^{-2} < &\aoz/ \Lambda^2 &  < 0.023~\mathrm{GeV}^{-2}, \\ 
 -0.029~\mathrm{GeV}^{-2} < &\acz/ \Lambda^2 & < 0.029~\mathrm{GeV}^{-2},  \\
 -0.020~\mathrm{GeV}^{-2} < &\aow/ \Lambda^2 & < 0.020~\mathrm{GeV}^{-2},  \\ 
 -0.052~\mathrm{GeV}^{-2} < &\acw/ \Lambda^2 & < 0.037~\mathrm{GeV}^{-2},
\end{eqnarray*}
where $\Lambda$ is the energy scale of the new physics. 
Limits found when allowing two or more parameters to vary are also presented.

\end{abstract}
 
\bigskip

\begin{center}
  {\large to be Submitted to Phys. Rev. D}
\end{center}
\end{titlepage} 

\begin{center}{\Large        The OPAL Collaboration
}\end{center}\bigskip
\begin{center}{
G.\thinspace Abbiendi$^{  2}$,
C.\thinspace Ainsley$^{  5}$,
P.F.\thinspace {\AA}kesson$^{  3,  y}$,
G.\thinspace Alexander$^{ 22}$,
J.\thinspace Allison$^{ 16}$,
P.\thinspace Amaral$^{  9}$, 
G.\thinspace Anagnostou$^{  1}$,
K.J.\thinspace Anderson$^{  9}$,
S.\thinspace Asai$^{ 23}$,
D.\thinspace Axen$^{ 27}$
% ...
}\end{center}\bigskip
\bigskip
$^{  1}$School of Physics and Astronomy, University of Birmingham,
Birmingham B15 2TT, UK
\newline
$^{  2}$Dipartimento di Fisica dell' Universit\`a di Bologna and INFN,
I-40126 Bologna, Italy
\newline
$^{  3}$Physikalisches Institut, Universit\"at Bonn,
D-53115 Bonn, Germany
\newline
% ...
$^{  5}$Cavendish Laboratory, Cambridge CB3 0HE, UK
\newline
% ...
$^{  9}$Enrico Fermi Institute and Department of Physics,
University of Chicago, Chicago IL 60637, USA
\newline
% ...
$^{ 16}$Department of Physics, Schuster Laboratory, The University,
Manchester M13 9PL, UK
\newline
% ...
$^{ 22}$Department of Physics and Astronomy, Tel Aviv University,
Tel Aviv 69978, Israel
\newline
$^{ 23}$International Centre for Elementary Particle Physics and
Department of Physics, University of Tokyo, Tokyo 113-0033, and
Kobe University, Kobe 657-8501, Japan
\newline
% ...
$^{ 27}$University of British Columbia, Department of Physics,
Vancouver BC V6T 1Z1, Canada
\newline
% ...
$^{  y}$ now at CERN
\newline
% ...
$^{  *}$ Deceased
\bigskip
 
 
\section{Introduction}
\label{sec:intro}

\begin{figure}[h]
\centerline{\epsfig{file=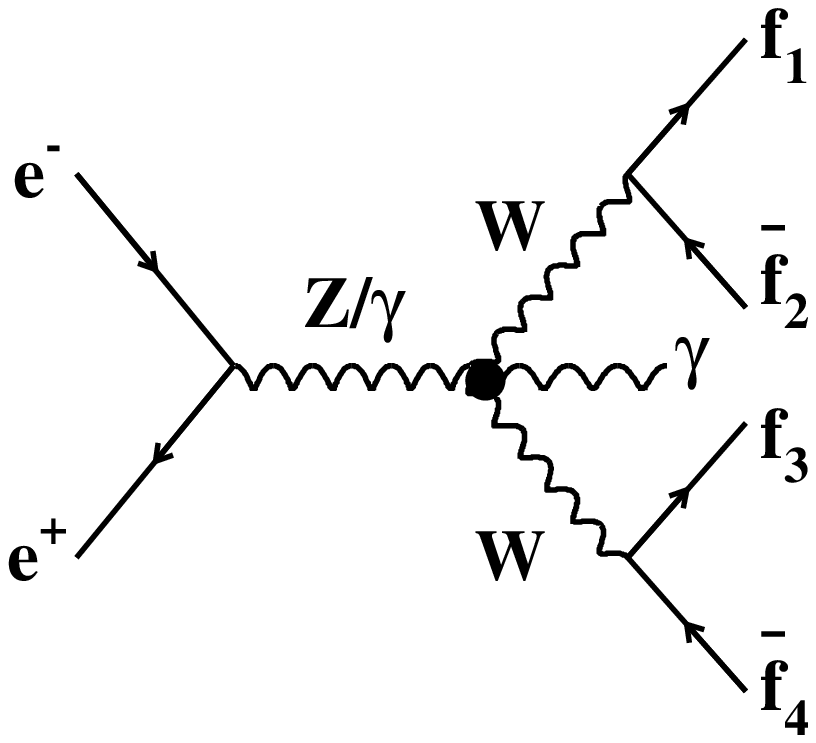, height=4.5cm}\hspace{5mm}\epsfig{file=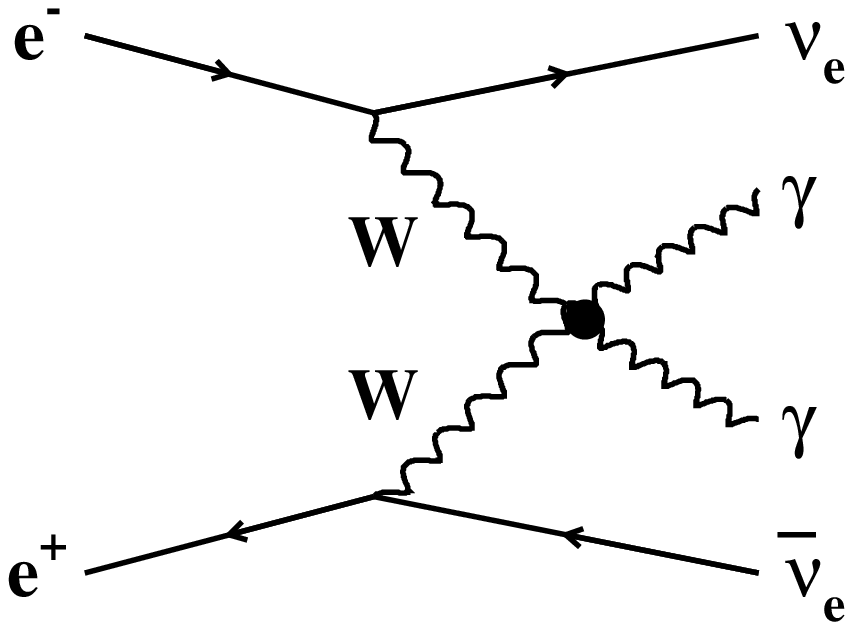, height=4.5cm}\hspace{5mm}\epsfig{file=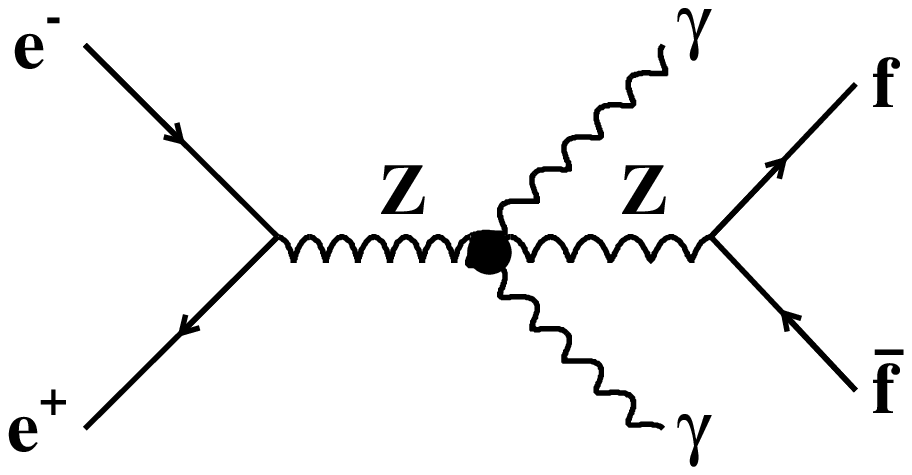, height=4.5cm}}
\caption{The diagrams  sensitive to possible 
anomalous quartic couplings in the $\Pep\Pem\to\WWg$, $\Pep\Pem\to\nngg$ and $\Pep\Pem\to\qqgg$ final states. }
\label{fig:qgcdiag}
\end{figure}

In the Standard Model (SM) self-interactions of the vector boson fields 
arise due to the   \mbox{$-\frac{1}{4}{\mathbf W_{\mu\nu}} \cdot {\bf W^{\mu\nu}}$} 
term in the electroweak Lagrangian.  
In addition to the tri-linear couplings, this 
term leads to quartic gauge couplings (QGCs) of the form 
WWWW, WWZZ, WW$\gamma\gamma$ and WWZ$\gamma$.  
The strength of the coupling at these vertices is specified by 
the $\mathrm{SU(2) \times U(1)}$ gauge invariant form of the electroweak 
sector. Studying processes to which these QGCs 
can contribute may therefore yield further confirmation of the non-Abelian structure of the SM or signal the presence of new physics at as yet unprobed energy scales.
At LEP energies it is 
only possible to probe quartic gauge couplings which produce at most two 
massive vector bosons in the final state. The processes at LEP which are 
sensitive to possible anomalous quartic gauge couplings (AQGCs) are shown 
in Figure~\ref{fig:qgcdiag}. 

The formalism for the extra genuine quartic terms relevant at LEP has been 
discussed widely in the 
literature~\cite{bib:boudjema,bib:stirling-abu,bib:stirling-qqgg,bib:stirling-wwg,bib:boudjema2,bib:RacoonQGC,bib:WRAP}. Genuine 
quartic terms refer to those that are not associated with any tri-linear 
couplings, which are already constrained by analyses using the  
$\Pep\Pem\to\WW$ process. In the parametrisation first introduced 
in~\cite{bib:boudjema} the two lowest dimension terms that give rise to 
quartic couplings involving at least two photons are:
\begin{eqnarray*}
\mathcal{L}_6^0 & = & - \frac{e^2}{16} \frac{a_0}{\Lambda^2} F_{\mu\nu}F^{\mu\nu}
\vec{W}^{\alpha} \cdot \vec{W}_{\alpha}, 
\\
\mathcal{L}_6^{\mathrm{c}} & = & - \frac{e^2}{16} \frac{a_\mathrm{c}}{\Lambda^2} F_{\mu\alpha}F^{\mu\beta}
\vec{W}^{\alpha} \cdot \vec{W}_{\beta}, 
\end{eqnarray*}  
where $F^{\mu\nu}$ is the photon field strength tensor.
These are C and P conserving and are obtained by imposing 
local $\mathrm{U(1)}_{\mathrm{em}}$ gauge symmetry, 
whilst also requiring the global custodial 
$\mathrm{SU(2)}_{\mathrm{c}}$ symmetry that preserves the constraint that the 
electroweak parameter $\rho=1$. 
We note that the custodial $\mathrm{SU(2)}_{\mathrm{c}}$ field vector is
\[
\vec{W}_{\alpha} = \left( \begin{array}{c} 
                           \frac{1}{\surd 2}(W_{\alpha}^+ + W_{\alpha}^-) \\
                           \frac{i}{\surd 2}(W_{\alpha}^+ - W_{\alpha}^-) \\
                           Z_{\alpha} / \cos\theta_W 
                           \end{array} \right) 
\]
and identifying
\[
\vec{W}_{\alpha} \cdot \vec{W}_{\beta} \to 2(W_{\alpha}^{+} W_{\beta}^{-} + \frac{1}{2\cos^2\theta_W} Z_{\alpha} Z_{\beta})
\]
yields, in terms of the physical fields, $W_{\alpha}^{+}$, $W_{\alpha}^{-}$ and $Z_\alpha$,
\begin{eqnarray*}
\mathcal{L}_6^0 & = & - \frac{e^2}{8} \frac{a_0^{\mathrm{W}}}{\Lambda^2} F_{\mu\nu}F^{\mu\nu} {W}^{+\alpha}{W}^-_{\alpha} 
- \frac{e^2}{16\cos^2\theta_W} \frac{a_0^{\mathrm{Z}}}{\Lambda^2} F_{\mu\nu}F^{\mu\nu} {Z}^{\alpha}{Z}_{\alpha},
\\
\mathcal{L}_6^{\mathrm{c}} & = & - \frac{e^2}{16} \frac{a_{\mathrm{c}}^{\mathrm{W}}}{\Lambda^2} F_{\mu\alpha}F^{\mu\beta} ({W}^{+\alpha}{W}^-_{\beta}+ 
W^{-\alpha}W^+_{\beta})- \frac{e^2}{16\cos^2\theta_W} \frac{a_{\mathrm{c}}^{\mathrm{Z}}}{\Lambda^2} F_{\mu\alpha}F^{\mu\beta} {Z}^{\alpha}{Z}_{\beta}.
\end{eqnarray*}  
Thus, both $\mathcal{L}_6^0$ and $\mathcal{L}_6^{\mathrm{c}}$ generate $\WW\gamma\gamma$ and 
$\ZZ\gamma\gamma$ couplings, with the parameters $a_0$ and $a_\mathrm{c}$ now being 
distinguished for the W and Z vertices to comply with the more general treatment 
in~\cite{bib:boudjema2}. In all cases the strengths of the quartic couplings are 
proportional to $1/\Lambda^2$ where $\Lambda$ is interpreted as the 
energy scale of the new physics.

Limits on AQGCs from LEP data have been published by the OPAL and 
L3 collaborations\cite{bib:OPALwwgold,bib:L3qqgg,bib:L3wwg,bib:OPALwwg}. 
This paper describes limits on AQGCs obtained by OPAL from the processes 
$\epem\rightarrow\nngg$ and $\epem\rightarrow\qqgg$ from all data recorded 
above the $\Zzero$ pole. For both processes the dominant SM background arises
from initial-state radiation (ISR). 
The limits obtained from $\epem\rightarrow\nngg$ and $\epem\rightarrow\qqgg$  
are combined with the limits obtained by OPAL from the process 
$\epem\rightarrow\WWg$\cite{bib:OPALwwg}.

Since cross-sections for the $\qqgg$ final state have not
previously been measured explicitly by the OPAL collaboration at LEP2, 
these measurements are presented in this paper and are compared with the 
SM expectation.


\section{The OPAL Detector and Data Samples}

The \Opal\ detector included a 3.7 m diameter tracking volume within
a 0.435~T axial magnetic field. The tracking detectors included a silicon
micro-vertex detector, a high precision gas vertex detector and a large
volume gas jet chamber. The tracking acceptance corresponds to approximately
$|\cos\theta|<0.95$ (for the track quality cuts used in this 
study)\footnote{The \Opal\ right-handed
                         coordinate system is defined such that the
                         origin is at the centre of the detector and the
                         $z$ axis points along the direction of the $e^-$
                         beam; $\theta$ is the polar angle with respect to 
                         the $z$ axis.}.
Lying outside the solenoid, 
the electromagnetic calorimeter (ECAL) 
consisted of $11\,704$ lead glass blocks
having  full acceptance in the range $|\cos\theta|<0.98$ and a relative
energy resolution of approximately $6\,\%$ for 10~GeV photons.
The hadron calorimeter consisted of the magnet return yoke 
instrumented with streamer tubes. Muon chambers outside the 
hadronic calorimeter provided muon identification in the range
$|\cos\theta|<0.98$.  A detailed description of the \Opal\ detector can be
found in \cite{bib:detector}.  

From 1995 to 2000 the LEP centre-of-mass energy was increased in 
several steps from 130 to 209~GeV. 
For the analysis of the  $\qqgg$ channel, this entire data sample is used, 
corresponding to 712~\ipb. The $\nngg$ analysis is restricted to 
652~\ipb\ of data recorded above 180~\GeV. 
The integrated luminosities at each centre-of-mass energy for the $\nngg$ analysis 
are lower than those for the 
$\qqgg$ analysis due to tighter requirements on the operational status of the
detector components.


\section{Monte Carlo Models}

A number of Monte Carlo (MC) samples, all including a  full simulation\cite{bib:GOPAL} 
of the \Opal\ detector, are used to simulate the SM signal and background processes. 
For the $\nngg$ final state \NUNUGPV~\cite{bib:nunugpv} is used to model both 
the dominant SM doubly-radiative return process and the supplementary AQGC processes, with \KKFF\cite{bib:kk} being used as a cross-check on the SM expectations.
For the $\qqgg$ final state, the \KKFF~program is also used. 
For the background processes, the concurrent MC tandem\cite{bib:KandY} of
\KoralW\ and YFSWW is used to simulate the background from four-fermion final
states with fermion flavour consistent with being from $\WW$ final states.
The \KoralW\ program\cite{bib:KoralW} is used to simulate the background
from four-fermion final states which are incompatible with
coming from the decays of two W-bosons ({\em e.g.} $\epem\rightarrow\qq\mpmm$). 
For both signal and background processes JETSET\cite{bib:Jetset} is used to model the 
fragmentation and hadronisation of final state quarks.   
The two-fermion background process
$\epem\rightarrow\Zzero/\gamma\rightarrow\tptm$ is 
simulated using \KKFF. The background in the
$\qqgg$ event selection from multi-peripheral two-photon diagrams
is negligible. The WRAP program\cite{bib:WRAP} is used to 
determine the effects of AQGCs in the $\qqgg$ channel.


\section{\boldmath The $\nngg$ Final State}
\label{sec:nngg}

\subsection{\boldmath $\nngg$ Event Selection}

The selection proceeds in two stages: 

\bigskip
\noindent
 {\underline{\bf Acoplanar photon pair selection:}} 
This event selection employs standard criteria described in detail elsewhere~\cite{bib:photon-pr, bib:photon-pr2}. 
Candidate events must meet the kinematic requirement of there being at least two photons, 
either both with energy $\Eg>0.05\Ebeam$ and polar angle $\theta_{\gamma}$ satisfying $\actg<0.966$, 
or one with $\Eg>0.05\Ebeam$, $\actg<0.966$ accompanied by a second with 
$\Eg>1.75$~GeV, $\actg<0.8$ that has an associated in-time time-of-flight detector signal.
Events with three final state photons $(\epem\to\nngg\gamma)$ are 
permitted, the subsequent selection criteria then being applied to the two photons with 
the highest reconstructed energies. The system consisting of the two highest energy 
photons must have a momentum transverse to the beam axis, $p_T^{\gamma\gamma}$, 
satisfying $p_T^{\gamma\gamma} > 0.05\Ebeam$. Additional requirements are then made on 
the photon conversion consistency (charged track veto), the electromagnetic calorimeter 
cluster shape, the forward energy vetoes and the muon vetoes. 
The $\Pep\Pem\to \Pgg\Pgg(\Pgg)$ background is suppressed whilst retaining the events with missing energy by imposing further cuts on the energies and angles of the selected two or three photon system. 
These include the requirements that the total energy in the electromagnetic
calorimeter does not exceed $0.95\roots$ and also that the 
acoplanarity\footnote{The acoplanarity angle is defined as $\pi$ minus the
opening angle between the two photons when projected onto a plane perpendicular
to the beam axis.} 
angle of the two highest energy photons be greater than 
$2.5^\circ$.

The efficiency for SM $\Pep\Pem\to\nngg({\gamma})$ events within the kinematic acceptance of the acoplanar photon pair selection is approximately 66\,\%~\cite{bib:photon-pr2}. The expected background contribution from processes other than $\Pep\Pem\to\nu\overline{\nu}\Pgg\Pgg(\Pgg)$ is less than 1\,\%~\cite{bib:photon-pr, bib:photon-pr2}. 

\bigskip
\noindent
{\underline{\bf Suppression of Standard Model background:}} 
To suppress the SM contribution, principally the forward-peaked 
doubly-radiative return process, the following additional cuts are 
applied to the events passing the acoplanar photon pair selection:
\begin{itemize}
\item The two highest reconstructed photon energies, $\Egamo$ and $\Egamt$, must 
both be greater than 10~GeV. This cut has little effect on any AQGC contribution, 
which gives rise predominantly to photons of high energy, but does suppress 
the doubly-radiative return background.
\item $|\cosgo|<0.9$, $|\cosgt|<0.9$, where again the subscripts refer to the two 
photons with highest reconstructed energy. This requirement further suppresses the 
doubly-radiative return background, which is forward peaked 
as expected for initial-state radiation photons.
\end{itemize}        
These cuts were optimised on SM MC to yield the 
maximum sensitivity to the anomalous couplings. 

\subsection{\boldmath Sensitivity of $\epem\rightarrow\nngg$ to Anomalous QGCs}

Table~\ref{tab:nevts} lists the number of data events accepted by the $\nngg$ event 
selection compared to the SM expectation, binned by centre-of-mass energy.
There is excellent agreement between the predictions of 
\NUNUGPV\ and the \KKFF\ MC program~\cite{bib:kk} used as a cross-check. The SM 
predictions describe the data well. 

Approximately 4.0$-$4.7\,\% of real data events, depending on the centre-of-mass energy, are expected to fail the acoplanar selection due to the effects of random coincidental activity.  These rates have been evaluated from samples of random beam-crossing events collected throughout the data-taking periods. All quoted MC accepted cross-sections have been corrected for these unmodelled 
effects.

\begin{table}[h]
\vskip 5mm
\centering
\begin{tabular}{|c|c|c|c|c|}
\hline
$\sqrt{s}$ & $\int \mathcal{L}\,\mathrm{d}t$ & Data  & \multicolumn{2}{c|}{SM Expectation} \\
 $[$GeV] & [pb$^{-1}$] &  & \NUNUGPV\  & \KKFF\  \\ \hline
180$-$185            &   \phantom{0}53.9   & \phantom{0}0    &   2.5      &   2.5   \\
188$-$190            &       175.2         & 10              &   7.9      &   7.9   \\
191$-$192            &   \phantom{0}28.8   &\phantom{0}1     &  1.3       &   1.3   \\
195$-$196            &   \phantom{0}71.6   &\phantom{0}0     &  3.1       &  3.0    \\
199$-$201            &   \phantom{0}73.7   &  \phantom{0}3   &  3.0       &   2.9   \\
201$-$203            &   \phantom{0}36.7   &  \phantom{0}1   &  1.5       &  1.4    \\
203$-$209            &          210.6      &  \phantom{0}5   &  8.3       &  8.0    \\ \hline
~Total~          &     652\phantom{.0}               & 20              &  27.6\phantom{0}      &  27.0\phantom{0}   \\
\hline
\end{tabular}
\caption{Numbers of $\nngg$ events passing the event selection by centre-of-mass energy
compared to the SM expectations from both \kkff\ and \nunugpv. All MC accepted 
cross-sections have been corrected for efficiency losses due to random coincident detector hits.}
\label{tab:nevts}
\vskip 5mm
\end{table}

For the selected events, 
Figure~\ref{fig:rmdeg2} shows the distribution of the invariant mass recoiling 
against the photons, 
$\mrec$, and the distribution 
of the energy of the photon with the second highest 
reconstructed energy, $\Egamt$. 
In both cases the data are well described by the SM expectation.
Figure~\ref{fig:rmdeg2} also shows the effects of anomalous couplings on  
these distributions. For the recoil mass, increasing the coupling at the 
$\ZZ\gamma\gamma$ vertex increases the cross-section at the $\Zzero$ mass peak, 
whereas the effect of the $\WW\gamma\gamma$ vertex can mainly be seen in the low 
recoil mass region of the plot.  Similarly, the effects of the different quartic 
vertices can be distinguished in different regions of the $\Egamt$ distribution. 

Constraints on AQGCs are derived employing a maximum likelihood fit 
that uses bins in the $\mrec$ and $\Egamt$ distributions 
at each centre-of-mass energy. The ten bins are defined in 
Table~\ref{tab:bins}, together with the corresponding numbers of 
events observed and expected in the SM summed over centre-of-mass 
energies.  The choice of binning 
reflects the differing effects of the anomalous 
couplings on the different regions of the $\mrec$ and $\Egamt$ distributions
and was optimised on SM MC for maximum sensitivity to the coupling parameters, 
inclusive of systematic effects.

\begin{table}[h]
\centering
\begin{tabular}{|c|c|c|c|c|}
\hline
Bin Number & $\mrec$ \ $[$GeV$]$ & $\Egamt$ \ $[$GeV$]$ & Observed & Expected \\ \hline
       1   &   $<$ 60            &  $10 - 25$  &     0     &  \phantom{$<$2}0.1   \\   
       2   &   $<$ 60            &  $25 - 45$  &     0     &  \phantom{2}$<$0.1   \\
       3   &   $<$ 60            & $>$45       &     0     &  \phantom{$2$}$<$0.1 \\
       4   &  $60 - 80$\phantom{$0$} &  $10 - 25$  &     1     &  \phantom{$<$2}0.5   \\
       5   &  $60 - 80$\phantom{$0$} &  $25 - 45$  &     2     &  \phantom{$<$2}0.4   \\
       6   &  $60 - 80$\phantom{$0$} &   $>$45 &     0     &  \phantom{$<$2}0.1   \\
       7   &  $80 - 120$         &  $10 - 25$  &     5     &  \phantom{$<$}11.7   \\ 
       8   &  $80 - 120$         &  $25 - 45$  &     6     &  \phantom{$<$2}8.3   \\
       9   &  $80 - 120$         &   $>$45     &     1     &  \phantom{$<$2}0.8   \\
       10  &   $>$ 120           &    $>$10    &     5     & \phantom{$<$2}5.7    \\ \hline 
\multicolumn{3}{|c|}{Total}                    &    20     &  \phantom{$<$}27.6   \\
\hline
\end{tabular}
\caption{The binning of the likelihood function for the $\nngg$ events together 
with the corresponding numbers of events observed and expected in the SM.}
\label{tab:bins}
\end{table}

\subsubsection{\boldmath Systematic Uncertainties ($\nngg$)}

The systematic errors in this analysis are found to be small in comparison to the 
statistical error from the 20 selected data events. 

\bigskip
\noindent
 {\underline{\bf Experimental uncertainties:}} 
The main experimental systematic uncertainty arises from the accuracy of the modelling 
of the energy scale and resolution of the electromagnetic calorimeter.  The evaluation 
of this is based on a comparison of reconstructed 
events with two beam-energy photons 
in the final state $\mathrm{e}^+\mathrm{e}^- \to \gamma\gamma$ with 
those simulated 
in MC.  
Additional degradations in the resolution and scaling were then applied 
to the accepted SM cross-sections (both total and in the analysis bins) 
to evaluate the systematic uncertainties, 
separately for the barrel ($|\cos\theta_{\gamma}|<0.7$) and end-cap 
($0.7<|\cos\theta_{\gamma}|<0.9$) regions of the detector and for each year of 
data taking. These uncertainties result in relatively large
fractional  
systematic uncertainties for individual analysis bins (approximately 20\,\% for the
bins with  smallest cross-section, {\em i.e.} bins 2 and 3 of Table~\ref{tab:bins}) 
though these propagate through to small overall errors of less than 1\,\% on the total cross-sections. 
Possible biases in the measured photon angle were found to be negligible.

\bigskip
\noindent
 {\underline{\bf Theory shape uncertainty:}}
The shapes of the SM $\mrec$ and $\Egamt$ distributions from  \KKFF\ and 
\NUNUGPV\ have been compared in order to evaluate any possible theoretical uncertainty in 
the SM prediction. Again, the variations in the total cross-sections were small $(<4\,\%)$, but large fractional 
variations could be seen for bins 1$-$3 which were hardly populated by the statistics available from \KKFF.

\bigskip
\noindent
 {\underline{\bf Normalisation uncertainty:}
Other sources of systematic uncertainty have been considered and 
affect primarily the overall normalisation.  The uncertainty related
to the modelling of initial-state radiation (ISR) 
has been assessed by turning off ISR with finite $p_T$, leading to a $\pm$5\,\% normalisation
uncertainty.
The cross-sections for \NUNUGPV\ have been compared with the predictions of
B\'elanger {\it et al.}~\cite{bib:boudjema2} and the difference used to estimate
a normalisation systematic uncertainty
of $\pm$4\,\%. In addition, the luminosity error is 
$\pm$0.3\,\%. 
These errors are added in quadrature to give an estimate of the overall 
normalisation uncertainty of 6.4\,\% which is taken to be independent of energy.

\bigskip 
At all centre-of-mass energies and for any combination of the couplings, 
the available NUNUGPV MC statistics amounts to at least one thousand times the data statistics
and the related MC statistical error is negligible.
Similarly, due to the large sample sizes of random events analysed, 
the uncertainties on the corrections for losses due to coincidental random 
detector hits are less than 1\,\% and are neglected. 
The systematic error associated with the expected background contribution 
from processes other than $\Pep\Pem\to\nu\overline{\nu}\Pgg\Pgg(\Pgg)$ is 
also negligible.
    

\subsection{\boldmath Limits on Anomalous QGCs from $\epem\rightarrow\nngg$}

At each centre-of-mass energy, 15 samples of 2\,000 events with 
differing values of 
$\aow, \aoz, \acw$ and $\acz$ have been simulated.  The 
extra Lagrangian terms are linear in the anomalous couplings.
Consequently, the cross-section has a quadratic dependence and 
these 15 samples are sufficient to parametrise fully 
$\sigma(\aow, \aoz, \acw, \acz$).  The generated
events are reweighted using matrix element weights from NUNUGPV to obtain 
Monte Carlo samples corresponding
to any combination of the anomalous QGCs ($\aow, \aoz, \acw, \acz$).

For the $\nngg$ final state, fits for each of the AQGC parameters have been 
performed to the data by summing the likelihood curves obtained from 
the seven centre-of-mass energies considered.  The effects of systematic uncertainties
are included in the fits.
The fitted AQGCs are compatible with 
zero and the resulting 95\,\% confidence level (C.L.) 
intervals on the anomalous couplings varied individually are listed in 
Table~\ref{tab:limits}. These limits include the effects of systematic
uncertainties. The corresponding likelihood curves
are shown in Figures~\ref{fig:1dcomb}a-\ref{fig:1dcomb}d.  
The results of a fit allowing two AQGC parameters to vary simultaneously are shown in 
Figure~\ref{fig:2dcomb}, again with the two parameters not plotted fixed at zero. Since anomalous ZZ$\gamma\gamma$ and $\WW\gamma\gamma$ couplings affect different regions
of the invariant mass and second photon energy distributions, the limits on $\aow$ and $\aoz$
are largely uncorrelated. The same is true for the limits on $\acw$ and $\acz$.

\begin{table}[htbp]
\renewcommand{\arraystretch}{1.1}
\begin{center}
\begin{tabular}{|c|c|c|} \hline 
   Process  & Coupling & 95\,\% C.L. Limit \\ \hline
   $\nngg$  & $\aoz$  &  $-0.009\,\GeV^{-2} < {\aoz}/{\Lambda^2} < 0.026\,\GeV^{-2}$ \\
   $\nngg$  & $\acz$  &  $-0.034\,\GeV^{-2} < {\acz}/{\Lambda^2} < 0.039\,\GeV^{-2}$ \\
   $\nngg$  & $\aow$  &  $-0.040\,\GeV^{-2} < {\aow}/{\Lambda^2} < 0.037\,\GeV^{-2}$ \\
   $\nngg$  & $\acw$  &  $-0.114\,\GeV^{-2} < {\acw}/{\Lambda^2} < 0.103\,\GeV^{-2}$ \\ \hline
   $\qqgg$  & $\aoz$  &  $-0.012\,\GeV^{-2} < {\aoz}/{\Lambda^2} < 0.027\,\GeV^{-2}$ \\
   $\qqgg$  & $\acz$  &  $-0.036\,\GeV^{-2} < {\acz}/{\Lambda^2} < 0.034\,\GeV^{-2}$ \\ \hline
   $\WWg$   & $\aow$  &  $-0.020\,\GeV^{-2} < {\aow}/{\Lambda^2} < 0.020\,\GeV^{-2}$ \\
   $\WWg$   & $\acw$  &  $-0.053\,\GeV^{-2} < {\acw}/{\Lambda^2} < 0.037\,\GeV^{-2}$ \\ \hline
\end{tabular}
\end{center}
\caption{The 95\,\% C.L. limits on the anomalous QGCs from the OPAL LEP2 data from 
         the processes shown in Figure~\ref{fig:qgcdiag}. 
         The $\nngg$ and $\qqgg$ results are described in this paper. The
         limits from the process $\epem\rightarrow\WWg$ are described in 
         Reference~\cite{bib:OPALwwg}. All limits include systematic uncertainties and 
         correspond to the
         case where only the coupling in question is varied from zero. 
\label{tab:limits} }
\renewcommand{\arraystretch}{1.0}
\end{table}


\section{\boldmath The $\qqgg$ Final State}

\label{sec:qqggacc}

In the SM, photons in the 
process $\epem\rightarrow\qqgg$ are radiated from either the initial or 
final state fermions. Photons from ISR tend
to be produced along the beam direction. Photons from final state
radiation (FSR) tend to be produced almost collinear with the
quarks and are often lost within hadronic jets.   
For the measurement of the $\qqgg$ cross-section 
a theoretical acceptance is defined which is well matched to the
experimental sensitivity. The cross-section is
defined within a $\qq$ invariant mass region dominated by the
$\Zzero$ exchange diagrams.

The $\epem\rightarrow\qqgg$ cross-section 
measured in this paper corresponds to the following
acceptance with respect to the $\qqgg$ system: 
\begin{itemize}
  \item There must be at least two photons satisfying:
  \subitem {\em i})\ $\Egami  > 5$~GeV, where  $\Egami$ is the energy of photon $i$, 
  \subitem {\em ii})\ $| \cosgi |< 0.95$, where $ \thegami$ is the polar angle of 
        photon $i$,
  \subitem {\em iii})\ $\cosgqi < 0.90$, where $\thegqi$ is the angle between photon
        $i$ and the direction of the nearest quark. 
  \item $|\Mqq-\Mz|<3\Gz$.
\end{itemize}
The quantity $\Mqq$ is defined as the propagator mass, {\em i.e.} the invariant
mass of the $\qq$ system before FSR. Photons from FSR are not considered as signal
and interference between ISR and FSR is neglected.

\subsection{\boldmath $\qqgg$ Event Selection}

The selection of the $\qqgg$ events proceeds in three stages:

\bigskip
\noindent
 {\underline{\bf\boldmath $\epem\rightarrow\qq$ event selection:}} 
       $\epem\rightarrow\qq$ events are selected using the algorithm 
       described in~\cite{bib:lep2mh}. 

\bigskip
\noindent
  \underline{\bf Photon identification:}
    Photon candidates can be identified as either unassociated 
    electromagnetic calorimeter (ECAL) clusters or  
    photon conversions, following the procedure described
    in~\cite{bib:OPALwwg}. 
    Only photons with measured energy $\Egam>5$~GeV and polar
    angle $|\cosg| < 0.95$ are retained.  
    The remainder of the event is 
    forced into two jets using the Durham algorithm~\cite{bib:Durham}. 
    Finally, to
    reduce background from photons from the decays of
    neutral hadrons, {\em e.g.} $\pi^0$ and $\eta$ decays, the photons
    are required to be isolated from the reconstructed jets 
    by requiring $\cosgj < 0.9$, where
    $\thegj$ is the angle between the photon and the direction of
    the closest reconstructed jet. Photon candidates which fail this 
    isolation criterion are merged to the nearest jet and the jet energy 
    is recalculated. Events with two or more identified photons 
    satisfying the above requirements are retained for the analysis. 
    For photons within the MC generator level acceptance
    $\Egam>5$~GeV, $|\cosg|<0.95$ and $\cosgq<0.9$, the photon identification
    efficiency is about 88\,\%. The requirement of two identified
    photons therefore rejects approximately 23\,\% of the $\qqgg$ signal.
 
\bigskip
\noindent
  \underline{\bf Kinematic requirements:} 
    The reconstructed mass of the hadronic system, $\Mqq$, is 
    required to be consistent with $\Mz$. For about 90\,\% of the
    events $\Mqq$ is obtained from a kinematic fit which imposes
    the constraints of energy and momentum conservation. In the 
    first instance the fit assumes a four-body final state consisting
    of two jets and two photons. If the fit 
    probability is less than 0.01, the fit is performed allowing
    for an unobserved photon along the $\epem$ beam axis. For events 
    where this fit probability is also less than 0.01, the 
    hadronic mass is taken to be the recoil mass calculated from the 
    reconstructed momenta of the two photons. 
    The number of events with mass reconstructed in the three possible
    categories is consistent with MC expectation. 
    The reconstructed invariant mass spectrum before the cut on $\Mqq$
    is shown in Figure~\ref{fig:fig2}.  Events within the region 
    $75~\GeV<\Mqq<125~\GeV$ are considered $\qqgg$ candidates. 
    The cut on $\Mqq$ removes 47 events in the data compared to the SM 
    expectation of 58.6. Due to
    experimental resolution this mass window is 
    larger than that used in the kinematic definition of the cross-section.  
    Nevertheless, this cut rejects approximately 6\,\% of the $\qqgg$ events
    satisfying the signal definition. 

After applying the cut on $\Mqq$ a total of 176 events are identified 
in the data, consistent with the SM expectation of 191.0.
Figures~\ref{fig:fig3}a-\ref{fig:fig3}e show
the distributions of $\Egamo$, $\Egamt$, $|\cosgo|$, $|\cosgt|$ 
and  $\Egamo+\Egamt$ for selected events. 
Figure~\ref{fig:fig3}f shows the 
distribution of the maximum $|\cosg|$ of the two highest energy 
photons in the event. In each case the data are in good 
agreement with the SM expectation.  

\subsection{Cross-section Results}
 
The $\qqgg$ cross-section is determined within the above acceptance definition.
Cross-section values are obtained for
the seven different centre-of-mass energy ranges listed in Table~\ref{tab:qqggresults}.
The $\qqgg$ cross-section is calculated from
\begin{eqnarray*}
     \sigqqgg & = & \frac{(N_{\mathrm{obs}}- \nback)}{\effqqgg {L}},
 \label{eqn:sigWWg}
\end{eqnarray*}
where $N_{\mathrm{obs}}$  is the accepted number of events,
$\nback$ is the SM expected number of background events,
and ${L} = \int\mathcal{L}\mathrm{d}t$ is the integrated luminosity, given in 
Table \ref{tab:qqggresults}. The $\qqgg$ selection efficiency, $\effqqgg$, 
is evaluated using the \KKFF\ MC samples and includes feed-through from genuine $\qqgg$ 
events outside the signal acceptance (a contribution of approximately
12\,\%).

The numbers of events selected at each energy
are listed in Table \ref{tab:qqggresults} along with the quantities used to
calculate the cross-sections. Also shown are the derived cross-sections for 
the above signal acceptance.
The systematic uncertainties are described below.
The results are consistent with the SM expectation,
as shown in Figure~\ref{fig:fig4}. 
Averaging over all energies, and taking into account correlated systematic
uncertainties the ratio of the observed to expected cross-sections is
\begin{eqnarray*}
    R(\mathrm{data}/\mathrm{SM}) = 0.92 \pm 0.07 \pm 0.04, 
\end{eqnarray*}
where the errors represent the statistical and systematic uncertainties
respectively.

\begin{table}[htbp]
\renewcommand{\arraystretch}{1}
\begin{center}
\begin{tabular}{|c|c|c|c|c|c|c|c|} \hline 
 $\roots$ & $<\roots>$ & $\int\mathcal{L}\mathrm{d}t$ & $\effqqgg$ & $\nback$   & $N_{\mathrm{obs}}$ & $\sigqqgg$ & $\sigqqgg(\mathrm{SM})$  \\ 
 $[$GeV$]$ & $[$GeV$]$ &  $[$pb$^{-1}$$]$ & $[$\%$]$ &   &  & $[$fb$]$ & $[$fb$]$  \\ \hline
130.0$-$137.0  &  133.0    &   10.6 &  $76.2\pm4.0$ & $1.1\pm0.3$  &   8 & $848\pm350\pm57$ & 738 \\  
160.0$-$173.0  &  166.9    &   20.3 &  $79.4\pm3.2$ & $1.0\pm0.2$  &   5 & $247\pm139\pm17$ & 412 \\   
180.0$-$185.0  &   182.7   &   57.2 &  $77.5\pm3.1$ & $2.7\pm0.5$  &  10 & $164\pm 71\pm13$ & 333 \\  
188.0$-$189.0  &  188.6    &  183.1 &  $77.7\pm2.9$ & $9.5\pm1.6$  &  53 & $305\pm 51\pm16$ & 309 \\  
 191.0$-$196.0 &  194.4    &  105.7 &  $77.4\pm2.9$ & $4.3\pm0.7$  &  25 & $254\pm 61\pm13$ & 288 \\  
199.0$-$204.0  & 200.2     &  114.1 &  $78.4\pm2.9$ & $3.0\pm0.6$  &  26 & $257\pm 57\pm12$ & 270 \\  
204.0$-$209.0  & 205.9     &  220.6 &  $76.0\pm2.9$ & $7.2\pm1.3$  &  49 & $250\pm 47\pm12$ & 257 \\  \hline
\end{tabular}
\end{center}
\caption{Selected $\qqgg$ events and
        cross-section results for the seven different 
       $\roots$ ranges used in the analysis.
        The $\roots$ range,  the mean luminosity weighted 
        value of $\roots$ and the corresponding integrated luminosity, 
        $\int\mathcal{L}\mathrm{d}t$, are listed. 
 For the measured cross-sections, the uncertainties are respectively 
       statistical and systematic. The uncertainties on the efficiencies and backgrounds  
       are the estimated systematic uncertainties including a contribution from finite
       MC statistics. Also shown is the SM expectation from \KKFF.
    \label{tab:qqggresults} }
\end{table}

\subsubsection{\boldmath \label{sec:sys} Systematic Uncertainties ($\qqgg$) }

The systematic uncertainties on the $\qqgg$ selection efficiency and
on the expected number of background events are 
estimated to be 2.7\,\% and approximately 20\,\% respectively.
The systematic uncertainties, described below, were obtained in the same 
manner as described in Ref.~\cite{bib:OPALwwg} where further 
details may be found. In addition the contributions to the
systematic uncertainties due to finite MC statistics are included in the
numbers listed in Table~\ref{tab:qqggresults}.

\bigskip
\noindent
\underline{\bf\boldmath Photon identification and isolation:}
A systematic uncertainty of 1\,\% is assigned to cover the
uncertainties in the simulation of the photon conversion rate 
and the accuracy of the simulation of the electromagnetic 
cluster shape\cite{bib:invHiggs}. 
The systematic error associated with the isolation requirements
depends on the accuracy of the MC simulation of the fragmentation
process in hadronic jets. This is verified in $\Zzero\rightarrow\qq$ events
recorded at $\roots\sim\Mz$ during $1998-2000$. For each selected
event, the inefficiency of the isolation requirements 
is determined for cones of varying half-angle 
defined around randomly orientated directions.
The inefficiency of the isolation cuts is 
parametrised as a function of the angle between the
cone and the nearest jet. For all cone half-angles the inefficiency
in the MC and data agree to better than 1\,\%; consequently
a 1\,\% systematic error is assigned. These two effects give a total 
uncertainty on the identification efficiency for a single photon of 1.4\,\%.
Since two photons are required in the analysis of $\qqgg$ this corresponds to 
an uncertainty in the $\qqgg$ efficiency of 2.8\,\%.

\bigskip
\noindent
\underline{\bf\boldmath Photon energy scale and resolution:}
A bias in the energy scale for photons (data relative to MC)
in the region of the energy cut, {\em i.e.} $\Egam\sim5$~GeV, would result
in a systematic bias in the $\qqgg$ cross-section measurement.
The uncertainty on the ECAL energy scale for photons in this region is
estimated by examining the invariant mass distribution of pairs of
photons from $\pi^0$ decays in 
$\epem\rightarrow\qq$ events recorded at $\roots\sim\Mz$ during 1998$-$2000
and  $\epem\rightarrow\qq(\gamma)$ events recorded at $\roots>180$~GeV.
As a result a 4\,\% systematic
uncertainty on the ECAL energy scale in the region of $\Egam\sim5$~GeV
is assigned. The resulting systematic uncertainty on the 
$\qqgg$ cross-section is 1.5\,\%.  

The systematic error from the uncertainty in the 
ECAL energy resolution is obtained in a similar manner to that used for
the ECAL energy scale using the same $\pi^0$ sample. There is no evidence
for a statistically significant 
difference between the energy scales in data and MC. 
The statistical precision of the comparison, $\pm10\,\%$, is used to 
assign the energy resolution uncertainty which, when propagated to the 
uncertainty on the $\qqgg$ cross-section,
yields a systematic error of $\pm0.6\,\%$.

\bigskip
\noindent
\underline{\bf\boldmath Photon angular acceptance:}
The systematic error associated with the acceptance requirement
of \mbox{$|\cosg|<0.95$} depends on the accuracy of the MC simulation 
of the angular reconstruction of ECAL clusters at the edge of the 
acceptance. By comparing the reconstructed polar angle of leptons
from different detectors (ECAL, tracking, muon chambers) 
in $\epem\rightarrow\epem$ and $\epem\rightarrow\mpmm$ events the ECAL
acceptance is known to $\pm3$~mrad. This uncertainty results in 
a 0.6\,\% uncertainty in the $\qqgg$ cross-section.

\bigskip
\noindent
\underline{\bf\boldmath Background uncertainties (\nback):}
The dominant source of background is from  
$\epem\rightarrow\Zgamma\rightarrow\qq\gamma$ where one of
the identified photons is from ISR and the other is 
associated with the hadronic jets. A photon associated with the 
hadronic jets may be either from FSR in the parton shower or from 
the decay  of a hadron ({\em e.g.} $\pi$ or $\eta$ decays). 
From the studies presented in~\cite{bib:OPALwwg} a 30\,\% systematic 
uncertainty on this background contribution is assumed. 
The systematic uncertainties on the small background
contributions from four-fermion events and from tau-pair events
are negligible.
An additional 0.8\,\% error is assigned to cover uncertainties in
the $\epem\rightarrow\qq$ selection.

\subsection{\boldmath Limits on Anomalous QGCs from $\epem\rightarrow\qqgg$}

The $\Pep\Pem \to \qqgg$ process is sensitive to the anomalous ZZ$\gamma\gamma$ vertex and the possible couplings $\aoz$, $\acz$. 
To set limits on these a binned maximum 
likelihood fit to the observed distribution of \Egamt\ is 
performed in 5 GeV bins. Fits are performed to the data for the seven 
separate energy ranges of Table~\ref{tab:qqggresults} 
and the resulting likelihood curves are summed. The effects of anomalous 
couplings are introduced by reweighting events generated with 
$\kkff$ using the ratio of anomalous QGC to SM matrix elements
obtained from the WRAP program\cite{bib:WRAP}. The resulting likelihood 
curves for one-dimensional fits to $\aoz$ and $\acz$ separately are 
shown in Figures~\ref{fig:1dcomb}a and \ref{fig:1dcomb}b.  
From these curves, 95\,\% C.L. upper limits on the anomalous 
couplings are obtained, shown in Table~\ref{tab:limits}. 
The limits include the effect of the experimental systematic errors 
and assume a 5\,\% theoretical uncertainty
(obtained by comparing the
predictions of \KKFF\ and WRAP over the centre-of-mass range considered in
this publication). 
The 95\,\% C.L. contour obtained 
from a simultaneous fit to $\aoz$ and $\acz$ is
shown in Figure~\ref{fig:2dcomb}a.


\section{\boldmath Combined Limits on Anomalous QGCs from the $\qqgg$, $\nngg$ and 
$\WWg$ processes}

The summed one-dimensional likelihood curves for the parameters $\aoz$ and 
$\acz$ from the $\qqgg$ and $\nngg$ final states are shown in
Figures~\ref{fig:1dcomb}a and ~\ref{fig:1dcomb}b. In this combination the small effect of correlated systematic
uncertainties between the two channels has been neglected\footnote{The correlated
component of the systematic uncertainty on the event selection efficiencies is 
estimated to be 2\,\%, dominated by correlated uncertainties from the photon energy scale and
photon angular acceptance.}. 
The corresponding combined 
95\,\% confidence level limits on possible anomalous contributions to the 
$\ZZ\gamma\gamma$ vertex are
\begin{eqnarray*}
 -0.007~\mathrm{GeV}^{-2} < &\aoz/ \Lambda^2 &  < 0.023~\mathrm{GeV}^{-2}, \\ 
 -0.029~\mathrm{GeV}^{-2} < &\acz/ \Lambda^2 & < 0.029~\mathrm{GeV}^{-2}.
\end{eqnarray*} 
When both  $\ZZ\gamma\gamma$ parameters are allowed to vary simultaneously the 
likelihood contours of Figure~\ref{fig:2dcomb}a are obtained.

The limits on possible anomalous contributions to the WW$\gamma\gamma$ vertex 
obtained here from the $\nngg$ channel are combined with the previous OPAL 
limits from the $\epem\rightarrow\WWg$ process\cite{bib:OPALwwg}. The resulting 
likelihood curves are shown in  Figures~\ref{fig:1dcomb}c and ~\ref{fig:1dcomb}d, 
again assuming the systematic 
uncertainties for the two channels are uncorrelated. The 
corresponding 95\,\% confidence level limits on anomalous contributions to
the $\WW\gamma\gamma$ vertex are:    
\begin{eqnarray*}
 -0.020~\mathrm{GeV}^{-2} < & \aow/ \Lambda^2 & < 0.020~\mathrm{GeV}^{-2}, \\ 
 -0.052~\mathrm{GeV}^{-2} < & \acw/ \Lambda^2 & < 0.037~\mathrm{GeV}^{-2}.
\end{eqnarray*}
The likelihood contours for these two parameters are shown in 
Figure~\ref{fig:2dcomb}b. 

In the literature the assumption that $a^\mathrm{Z}_i=a^\mathrm{W}_i$
has been made (see for example Ref.~\cite{bib:stirling-wwg}). The validity
of the linking of the $\WW\gamma\gamma$ and $\ZZ\gamma\gamma$ couplings
has been questioned in Ref.~\cite{bib:boudjema2}. For completeness, limits 
are presented for the case where $a^\mathrm{Z}_i=a^\mathrm{W}_i$ by combining
the one-dimensional likelihood curves from the $\nngg$, $\qqgg$ and $\WWg$ 
processes, shown in Figures~\ref{fig:1dcomb}e and ~\ref{fig:1dcomb}f. The combined likelihood yields 
the 95\,\% confidence level limits:
\begin{eqnarray*}
+0.002  \mathrm{~GeV}^{-2} < & \aov/ \Lambda^2 & < 0.019 \mathrm{~GeV}^{-2}, \\
-0.022  \mathrm{~GeV}^{-2} < & \acv/ \Lambda^2 & < 0.029 \mathrm{~GeV}^{-2}. 
\end{eqnarray*}
The corresponding two-dimensional fit is shown in Figure~\ref{fig:finalcomb}.


\section{Conclusion}

Event selections for the processes $\nngg$ and $\qqgg$ are presented. 
The selected $\qqgg$ events are used to measure the cross-section for the process
$\epem\rightarrow\qqgg$. Averaging over all energies, the ratio of the
observed $\epem\rightarrow\qqgg$ cross-section to the Standard Model expectation is
\begin{eqnarray*}
    \mathrm{R}(\mathrm{data}/\mathrm{SM}) = 0.92 \pm 0.07 \pm 0.04, 
\end{eqnarray*}
where the errors represent the statistical and systematic uncertainties
respectively.
The selected $\nngg$ and $\qqgg$ events are used to constrain possible anomalous 
$\WW\gamma\gamma$ and $\ZZ\gamma\gamma$ couplings. When these results are 
combined with previous OPAL results from the $\WWg$ final state the 95\,\% 
confidence level limits on the anomalous coupling parameters 
$\aoz$, $\acz$, $\aow$ and $\acw$ are found to be:
\begin{eqnarray*}
 -0.007~\mathrm{GeV}^{-2} < &\aoz/ \Lambda^2 & < 0.023~\mathrm{GeV}^{-2}, \\ 
 -0.029~\mathrm{GeV}^{-2} < &\acz/ \Lambda^2 & < 0.029~\mathrm{GeV}^{-2},  \\
 -0.020~\mathrm{GeV}^{-2} < &\aow/ \Lambda^2 & < 0.020~\mathrm{GeV}^{-2},  \\ 
 -0.052~\mathrm{GeV}^{-2} < &\acw/ \Lambda^2 & < 0.037~\mathrm{GeV}^{-2},
\end{eqnarray*}
where $\Lambda$ is the energy scale of the new physics. 
Limits allowing two or more parameters to vary are also presented.


\section{Acknowledgments}

We particularly wish to thank the SL Division for the efficient operation
of the LEP accelerator at all energies
 and for their close cooperation with
our experimental group.  In addition to the support staff at our own
institutions we are pleased to acknowledge the  \\
Department of Energy, USA, \\
National Science Foundation, USA, \\
Particle Physics and Astronomy Research Council, UK, \\
Natural Sciences and Engineering Research Council, Canada, \\
Israel Science Foundation, administered by the Israel
Academy of Science and Humanities, \\
Benoziyo Center for High Energy Physics,\\
Japanese Ministry of Education, Culture, Sports, Science and
Technology (MEXT) and a grant under the MEXT International
Science Research Program,\\
Japanese Society for the Promotion of Science (JSPS),\\
German Israeli Bi-national Science Foundation (GIF), \\
Bundesministerium f\"ur Bildung und Forschung, Germany, \\
National Research Council of Canada, \\
Hungarian Foundation for Scientific Research, OTKA T-038240, 
and T-042864,\\
The NWO/NATO Fund for Scientific Research, the Netherlands.\\


\newpage

\begin{figure}[htbp]
\centerline{\epsfig{file=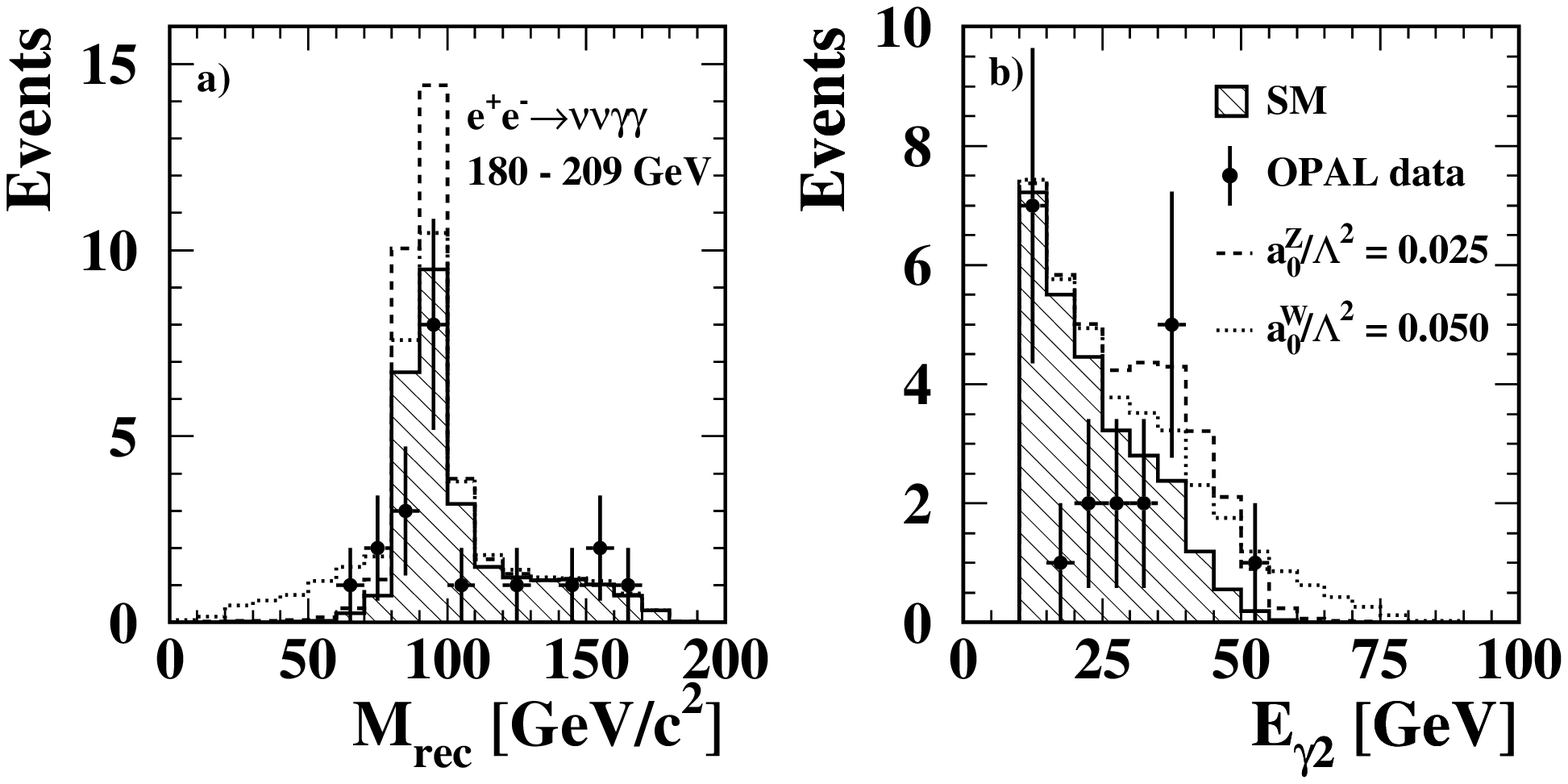, height=9.5cm}}
\caption{Distributions of a) $\mrec$ and b) $\Egamt$ for the accepted $\nngg$ events. 
The points show the $180-209$~GeV data and the histograms show the MC expectation. 
The hatched histogram represents the SM scenario whilst the expected distributions 
for possible $\ZZ\gamma\gamma$ and $\WW\gamma\gamma$ AQGC hypotheses are shown 
by the dashed and dotted lines, respectively.}
\label{fig:rmdeg2}
\end{figure}

\newpage
\begin{figure}[pt]
\centerline{\mbox{\epsfig{file=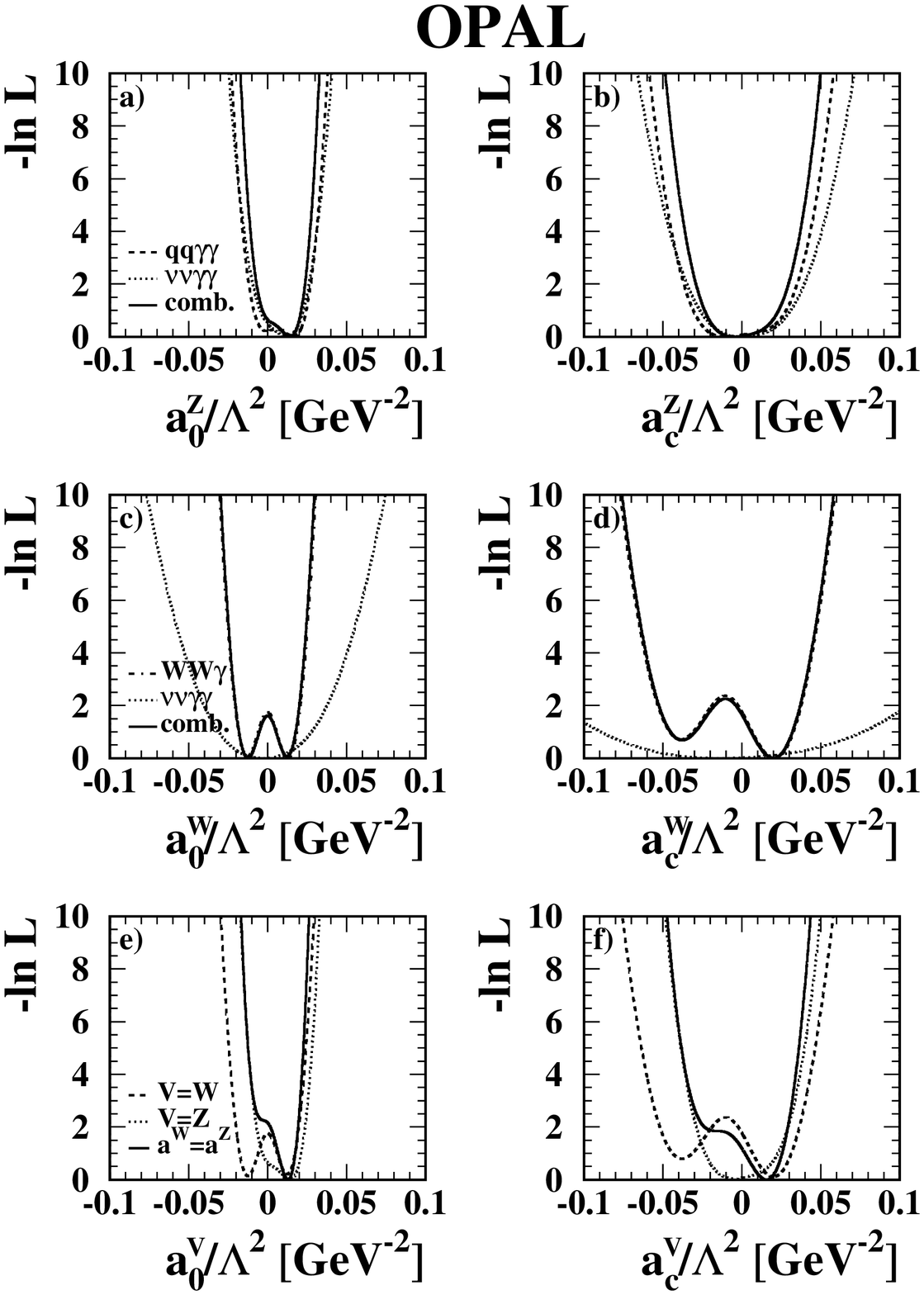, height=18cm}}}
\caption{Plots a) and b) show the one dimensional minus log likelihood curves for 
         $\aoz$ and $\acz$ from the 
         $\nngg$ channel (dotted line), the $\qqgg$ channel (dashed line), and the
         two channels combined (continuous line).
Plots c) and d) show the one dimensional likelihood curves for 
         $\aow$ and $\acw$ from the 
         $\nngg$ channel (dotted line), the $\WWg$ channel (dashed line), and the
         two channels combined (continuous line). 
         Figures e) and f) show the combined limits assuming $\aoz=\aow$ and 
         $\acz=\acw$. 
{{e)}} The one dimensional likelihood curve for $\aov=\aoz=\aow$ 
 (continuous line) with the contribution from the  
 $\aoz$ from the $\qqgg$ and $\nngg$ channels (dotted line) and from the
 limit on $\aow$ from the $\WWg$ and $\nngg$ channels (dashed line).
{{f)}} The one dimensional likelihood curve for $\acv=\acz=\acw$ 
 (continuous line) with the contribution from  
 $\acz$ from the $\qqgg$ and $\nngg$ channels (dotted line) and from the
 limit on $\acw$ from the $\WWg$ and $\nngg$ channels (dashed line).
         All likelihood curves include the effects of systematic uncertainties and
         correspond to the
         case where only the coupling in question is varied from zero.
}
\label{fig:1dcomb}
\end{figure}

\newpage
\begin{figure}[pb]
\centerline{\mbox{\epsfig{file=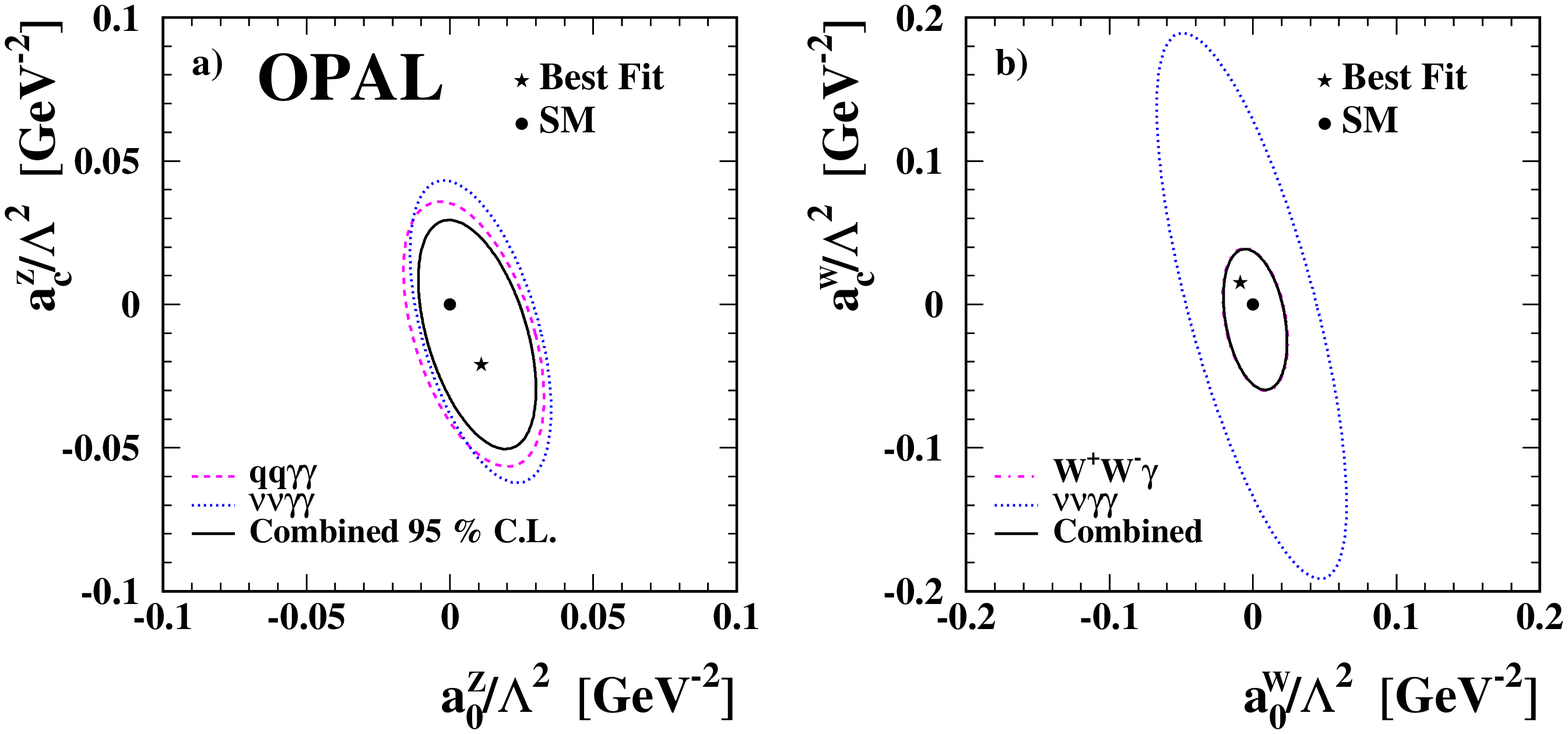, height=10cm}}}
\caption{ {a)} The 95\,\% confidence region in ($\aoz, \acz$) from the 
         $\nngg$ channel (dotted line), the $\qqgg$ channel (dashed line), and the
         two channels combined (continuous line). 
 {b)} The 95\,\% confidence region in ($\aow, \acw$) from the 
         $\nngg$ channel (dotted line), the $\WWg$ channel\cite{bib:OPALwwg} (dashed line), 
         and the
         two channels combined (continuous line). In {b)} the limits from
         the $\WWg$ channel dominate to such an extent that the limits from
         the $\WWg$ channel alone almost coincide with the combined limit.
         In both  {a)} and {b)} the position of the best fit (minimum
         of the $-\ln$L surface) is indicated by the star and the SM expectation at 
         $(0,0)$ is shown by the point.}
\label{fig:2dcomb}
\end{figure}

\newpage
\begin{figure}[htbp]
 \begin{center}
 \epsfxsize=\textwidth
 \epsffile{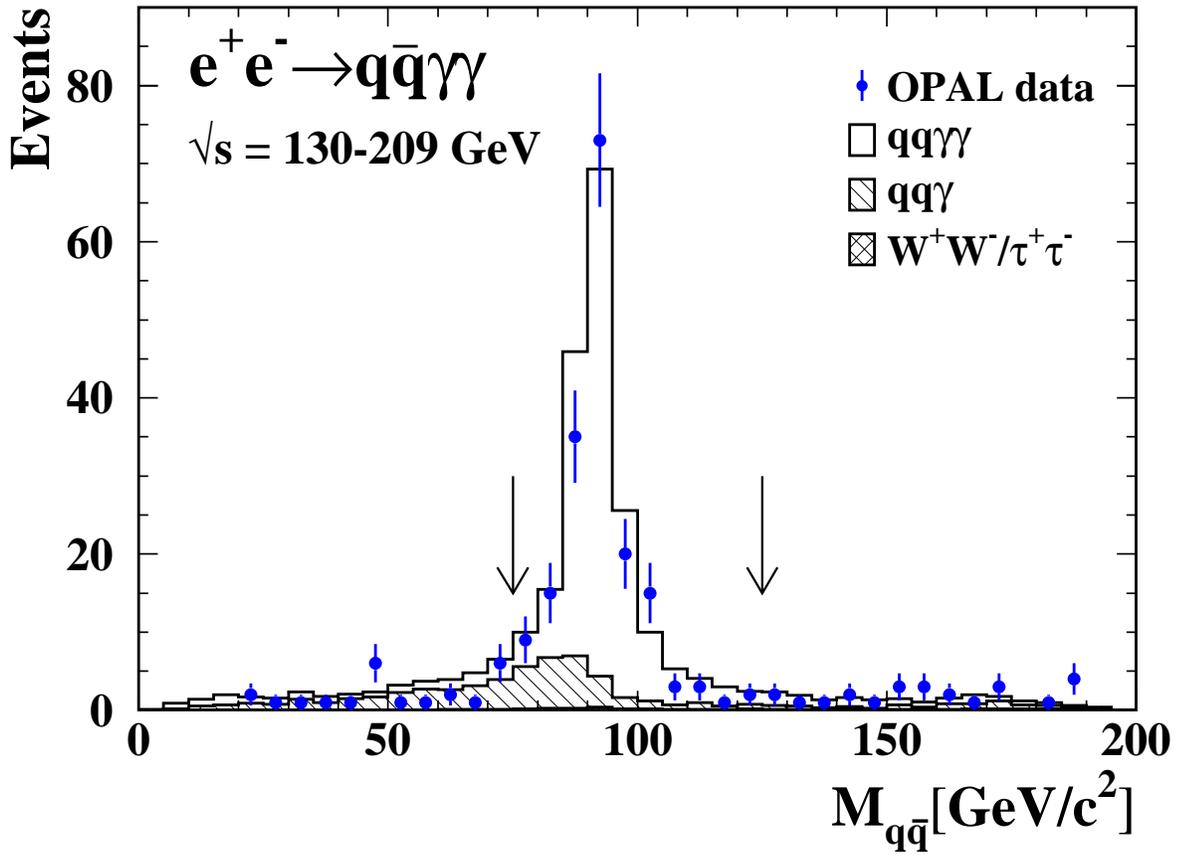}
 \caption{Invariant mass of the hadronic system, \Mqq, in selected
          $\qqgg$ events. The arrows indicate the cuts used to select
          the final $\qqgg$ sample. The singly hatched histogram 
          indicates the background from $\qqg$ events and the doubly
          hatched histogram (barely visible) indicates the small 
          four-fermion and tau-pair
          backgrounds.}
 \label{fig:fig2}
 \end{center}
\end{figure}

\newpage
\begin{figure}[htbp]
 \begin{center}
 \epsfxsize=16cm
 \epsffile{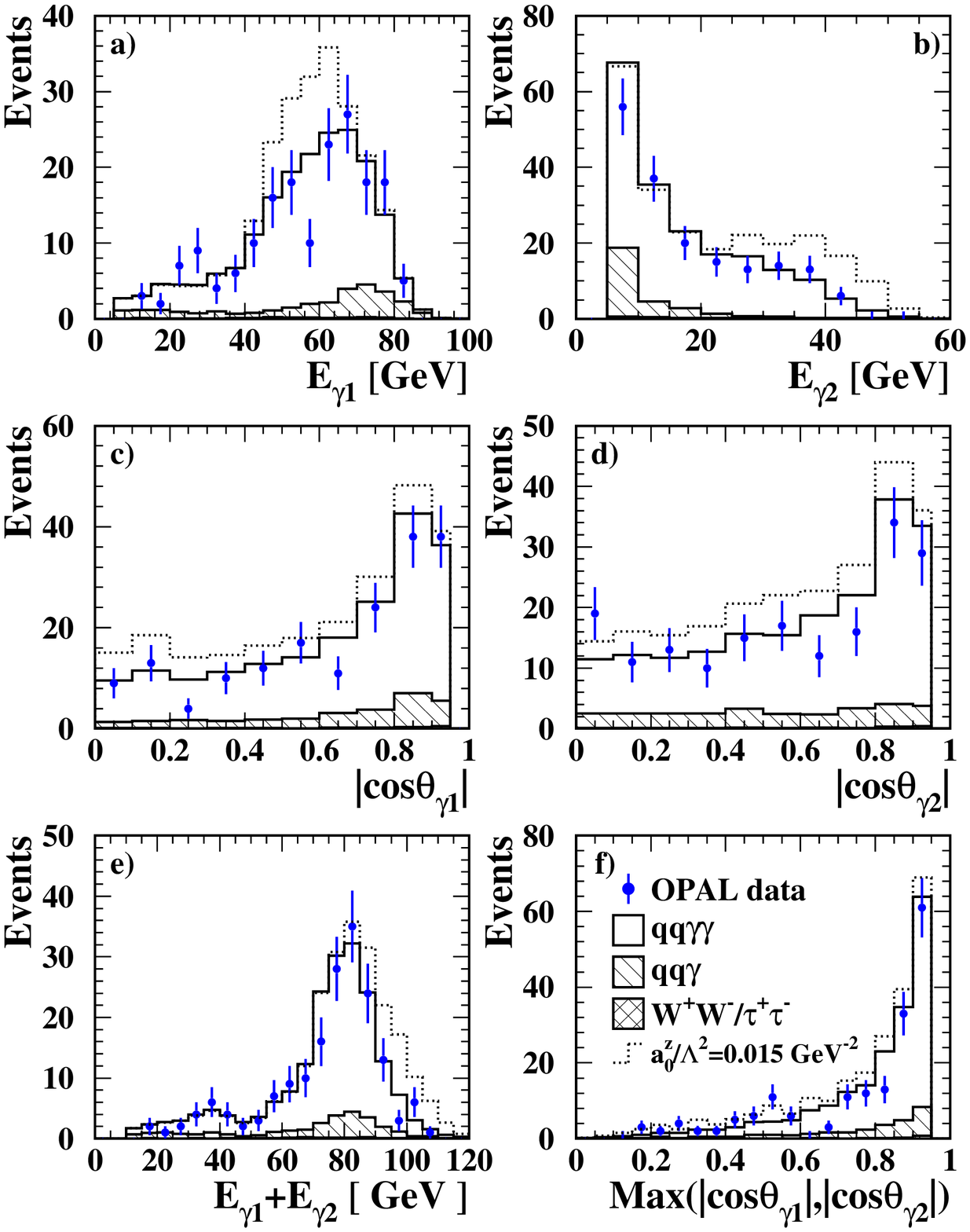}
 \caption{ Distributions of $\Egamo$, $\Egamt$, $|\cosgo|$, $|\cosgt|$,
           $\Egamo+\Egamt$ 
           and the maximum of $|\cosgo|$ and $|\cosgt|$ for selected
           $\qqgg$ events. The points show the $130-209$~GeV data
           and the histograms show the MC expectation.
           The singly hatched histogram 
           indicates the background from $\qqg$ events and the doubly
           hatched histogram indicates the four-fermion and tau-pair
           backgrounds. The expected distributions for 
           an anomalous QGC parametrised by 
           $\aoz/\Lambda^2=0.015~\GeV^{-2}$ are shown by the dotted 
           lines.}
 \label{fig:fig3}
 \end{center}
\end{figure}

\newpage
\begin{figure}[htbp]
 \begin{center}
 \epsfxsize=\textwidth
 \epsffile{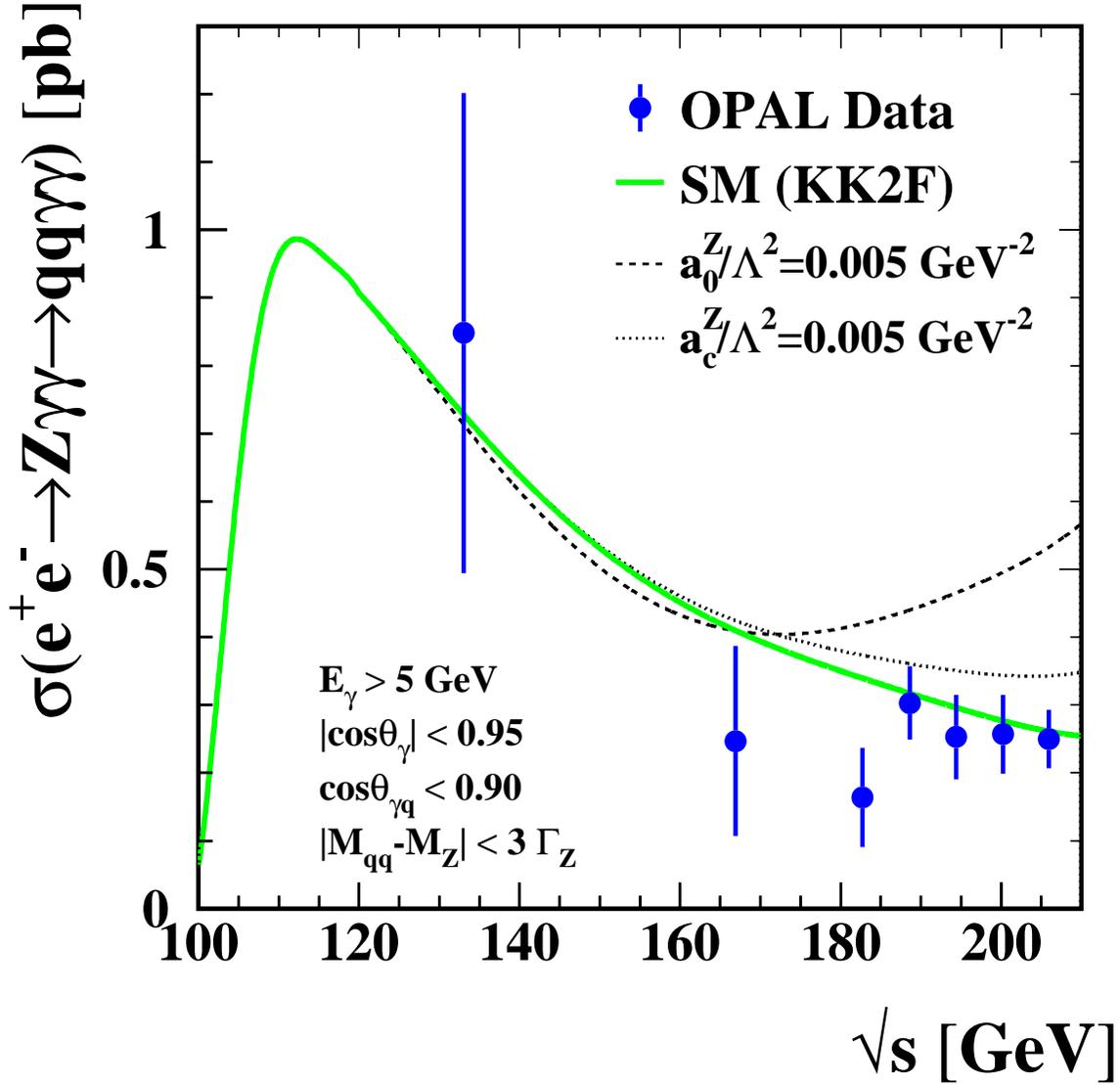}
 \caption{ Measured  $\epem\rightarrow\qqgg$ cross-section versus
          $\roots$. The cross-section corresponds to the 
          definition given in Section \ref{sec:qqggacc}.
          The SM prediction is obtained from 
          \KKFF\ (without contributions from FSR). The dashed and dotted curves show 
          the effects of anomalous QGCs on the cross-section.}
 \label{fig:fig4}
 \end{center}
\end{figure}

\newpage

\begin{figure}[pb]
\centerline{\epsfig{file=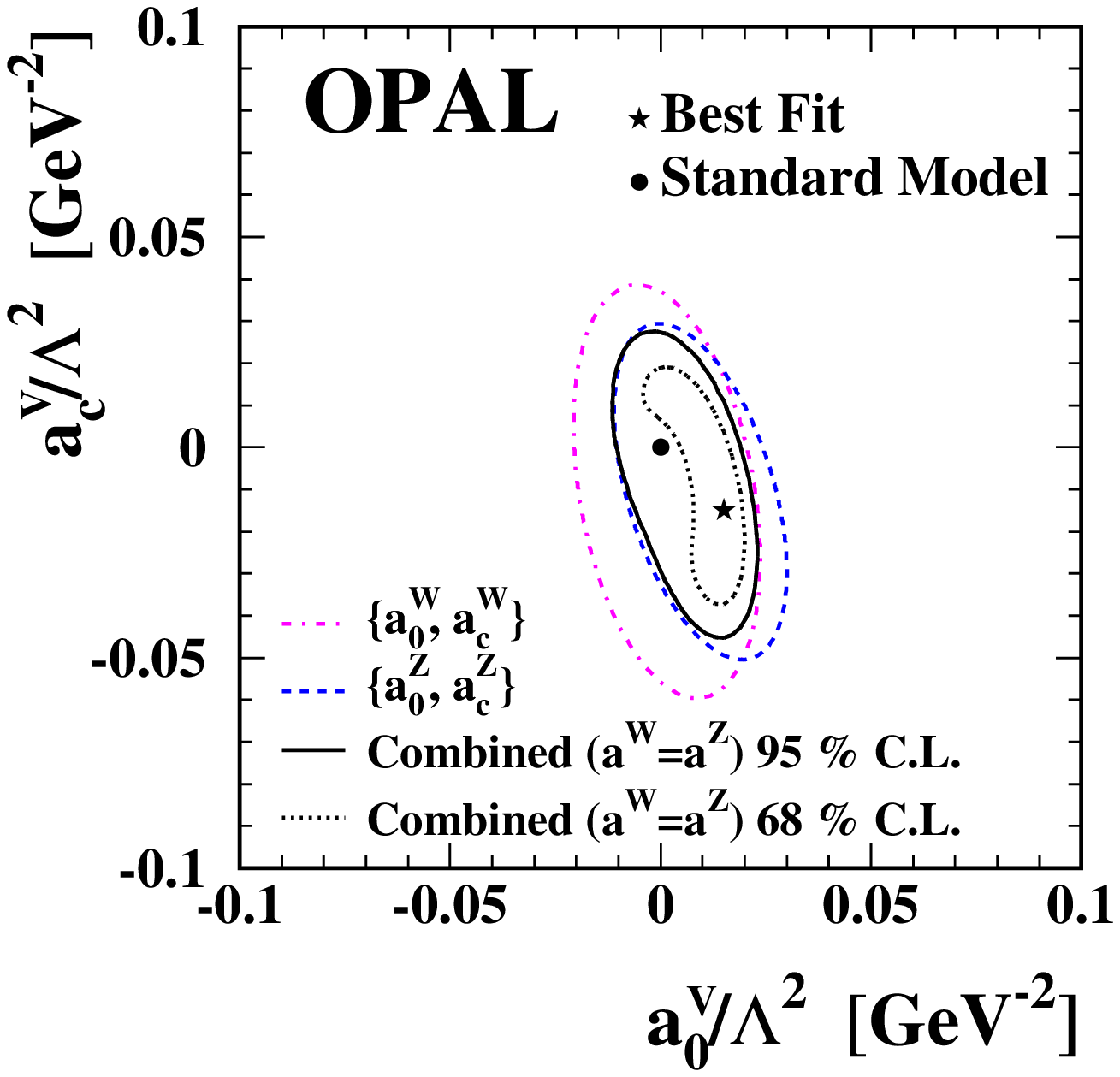, height=15cm}}
\caption{The 95\,\% confidence region in ($\aov, \acv$) 
          assuming $\aoz=\aow$ and $\acz=\acw$ (continuous line). 
          Also shown is the 68\,\% confidence region (dotted line). 
          The separate limits on ${\aoz,\acz}$ from the $\qqgg$ and $\nngg$  
          channels (dashed line) and from the limits on ${\aow,\acw}$ 
	  from the $\WWg$ and $\nngg$ channels (dot-dashed line) are also
          shown.
The position of the best fit (minimum
         of the $-\ln$L surface) is indicated by the star. The SM expectation at 
         $(0,0)$ is shown by the point.}
\label{fig:finalcomb}
\end{figure}

\end{document}